\documentstyle[epsfig]{mn}

\newif\ifAMStwofonts
\ifoldfss
  \newcommand{\rmn}[1] {{\rm #1}}

  \ifCUPmtlplainloaded \else
    \NewTextAlphabet{textbfit} {cmbxti10} {}
    \NewTextAlphabet{textbfss} {cmssbx10} {}
    \NewMathAlphabet{mathbfit} {cmbxti10} {} % for math mode
    \NewMathAlphabet{mathbfss} {cmssbx10} {} %  "   "    "
  \fi
  \ifAMStwofonts
    \ifCUPmtlplainloaded \else
      \NewSymbolFont{upmath} {eurm10}
      \NewSymbolFont{AMSa} {msam10}
      \NewMathSymbol{\upi}     {0}{upmath}{19}
      \NewMathSymbol{\umu}     {0}{upmath}{16}
      \NewMathSymbol{\upartial}{0}{upmath}{40}
      \NewMathSymbol{\leqslant}{3}{AMSa}{36}
      \NewMathSymbol{\geqslant}{3}{AMSa}{3E}

      \let\geq=\geqslant 
    \fi
  \fi
\fi % End of OFSS

\ifnfssone
  \newmathalphabet{\mathit}
  \addtoversion{normal}{\mathit}{cmr}{m}{it}
  \addtoversion{bold}{\mathit}{cmr}{bx}{it}
  \newcommand{\rmn}[1] {\mathrm{#1}}

  \newmathalphabet{\mathbfit} % math mode version of \textbfit{..}
  \addtoversion{normal}{\mathbfit}{cmr}{bx}{it}
  \addtoversion{bold}{\mathbfit}{cmr}{bx}{it}
  \newmathalphabet{\mathbfss} % math mode version of \textbfss{..}
  \addtoversion{normal}{\mathbfss}{cmss}{bx}{n}
  \addtoversion{bold}{\mathbfss}{cmss}{bx}{n}
  \ifAMStwofonts
    \ifCUPmtlplainloaded \else
      \UseAMStwoboldmath
      \makeatletter
      \new@mathgroup\upmath@group
      \define@mathgroup\mv@normal\upmath@group{eur}{m}{n}
      \define@mathgroup\mv@bold\upmath@group{eur}{b}{n}
      \edef\UPM{\hexnumber\upmath@group}
      \new@mathgroup\amsa@group
      \define@mathgroup\mv@normal\amsa@group{msa}{m}{n}
      \define@mathgroup\mv@bold\amsa@group{msa}{m}{n}
      \edef\AMSa{\hexnumber\amsa@group}
      \makeatother
      \mathchardef\upi="0\UPM19
      \mathchardef\umu="0\UPM16
      \mathchardef\upartial="0\UPM40
      \mathchardef\leqslant="3\AMSa36
      \mathchardef\geqslant="3\AMSa3E

      \let\geq=\geqslant 
    \fi
  \fi
\fi % End of NFSS release 1

\ifnfsstwo
  \newcommand{\rmn}[1] {\mathrm{#1}}

  \DeclareMathAlphabet{\mathbfit}{OT1}{cmr}{bx}{it}
  \SetMathAlphabet\mathbfit{bold}{OT1}{cmr}{bx}{it}
  \DeclareMathAlphabet{\mathbfss}{OT1}{cmss}{bx}{n}
  \SetMathAlphabet\mathbfss{bold}{OT1}{cmss}{bx}{n}
  \ifAMStwofonts
    \ifCUPmtlplainloaded \else
      \DeclareSymbolFont{UPM}{U}{eur}{m}{n}
      \SetSymbolFont{UPM}{bold}{U}{eur}{b}{n}
      \DeclareSymbolFont{AMSa}{U}{msa}{m}{n}
      \DeclareMathSymbol{\upi}{0}{UPM}{"19}
      \DeclareMathSymbol{\umu}{0}{UPM}{"16}
      \DeclareMathSymbol{\upartial}{0}{UPM}{"40}
      \DeclareMathSymbol{\leqslant}{3}{AMSa}{"36}
      \DeclareMathSymbol{\geqslant}{3}{AMSa}{"3E}

      \let\geq=\geqslant 
    \fi
  \fi
\fi % End of NFSS release 2

\ifCUPmtlplainloaded \else
  \ifAMStwofonts \else % If no AMS fonts
    \def\upi{\pi}
    \def\umu{\mu}
    \def\upartial{\partial}
  \fi
\fi

\title{Occurrence of Metal-free Galaxies in the Early Universe}
\author[J.L. Johnson, T.H. Greif, and V. Bromm]
       {Jarrett L. Johnson$^1$\thanks{E-mail: jljohnson@astro.as.utexas.edu} , Thomas H. Greif $^2$$^,$$^3$, and Volker Bromm$^1$ \\
 $^1$Department of Astronomy, University of Texas, Austin, TX 78712, USA \\
 $^2$Institut f\"{u}r Theoretische Astrophysik, Universit\"{a}t Heidelberg, Albert-Ueberle-Strasse~2, 69120 Heidelberg, Germany \\
 $^3$Fellow of the International Max Planck Research School for Astronomy and Cosmic Physics at the University of Heidelberg
\\}

\pubyear{2007}

\begin{document}

\maketitle
\topmargin-1cm

\label{firstpage}

\begin{abstract}
The character of the first galaxies at redshifts $z\ga 10$ strongly depends on
their level of pre-enrichment, which is in turn determined by the rate of
primordial star formation prior to their assembly. In order for the first
galaxies to remain metal-free, star formation in minihaloes must be highly
suppressed, most likely by H$_2$-dissociating Lyman-Werner (LW) radiation.
We show that the build-up of such a strong LW background is hindered
by two effects. Firstly, the level of the LW background is self-regulated, 
being produced by the Population~III (Pop~III) star formation which it, in turn, suppresses.  
Secondly, the high opacity to LW photons which is built up in the relic H~II 
regions left by the first stars acts to diminish the global LW background.
Accounting for a
self-regulated LW background,  we estimate a lower limit for the rate of Pop III star formation in minihaloes at $z$ $\ga$ 15.  
Further, we simulate the formation of a 'first
galaxy' with virial temperature $T_{\rm{vir}}\ga 10^{4}~\rm{K}$ and total mass $\ga$ 10$^8$ M$_{\odot}$ at $z\ga 10$,
and find that complete suppression of previous Pop~III star formation is unlikely, with stars of $\ga 100~\rm{M_{\odot}}$ (Pop III.1)
and $\ga 10~\rm{M_{\odot}}$ (Pop~III.2) likely forming. Finally, we discuss
the implications of these results for the nature of the first galaxies, which
may be observed by future missions such as the {\it James Webb Space
Telescope}.
\end{abstract}


\section {Introduction}
How did the first galaxies in the Universe form?  In the hierarchical picture of structure formation, the first galaxies, with masses of $\ga$ 10$^8$ M$_{\odot}$, were built up from the mergers of smaller dark matter (DM) minihaloes, with virial temperatures $\la$ 10$^4$ K, in which the first Pop III star formation may have occurred (e.g. Bromm et al. 1999, 2002; Abel et al. 2002).  The feedback effects from the first stars forming in minihaloes, likely having masses of the order of 100 M$_{\odot}$ (e.g. Tan \& McKee 2004; Yoshida et al. 2006), may thus have established the properties of the gas from which the first galaxies formed (see e.g. Ciardi \& Ferrara 2005).  During their brief lives of $\sim$ 3 Myr, very massive Pop III stars emit copious UV radiation which can ionize the primordial gas and destroy H$_2$ molecules (e.g. Schaerer 2002), and upon their deaths they may explode as powerful supernovae, expelling the first heavy elements into their surroundings (Mori et al. 2002; Bromm et al. 2003; Kitayama et al. 2005; Greif et al. 2007).  

%VB: small change of wording in this paragraph (too many "from which").
The nature of the first galaxies may largely be determined by the metallicity of the gas from which they form, as the character of star formation is predicted to transition from a massive Pop III initial mass function (IMF) to a low-mass dominated IMF when the primordial gas has been enriched to a critical metallicity $Z_{\rm crit}$ (e.g. Bromm et al. 2001a; Schneider et al. 2003; Santoro \& Shull 2006; Frebel et al. 2007; but see Jappsen et al. 2007). In turn, the metallicity of the protogalactic gas is dependent on the preceding Pop~III star formation that takes place in minihaloes. While the supernova explosions of Pop III stars formed during the assembly of the first galaxies could drive the metallicity of these systems to super-critical levels (see e.g. Greif et al. 2007; Karlsson et al. 2008), Pop III star formation in minihaloes could, in principle, be largely suppressed due to radiative feedback effects (e.g. Haiman et al. 2000; Mackey et al. 2003). 

Much recent work has been devoted to studying the effects of both ionizing and molecule-dissociating, Lyman-Werner (LW), radiation emitted from local star forming regions on Pop III star formation in minihaloes (Glover \& Brand 2001; Shapiro et al. 2004; Alvarez et al. 2006; Abel et al. 2006; Susa \& Umemura 2006; Ahn \& Shapiro 2007; Yoshida et al. 2007; Johnson et al. 2007; Whalen et al. 2007), with results generally suggesting that local intermittent LW feedback does little to delay star formation, while ionizing radiation only delays star formation in minihaloes with relatively low-density gas that may not efficiently form stars even in the absence of radiative effects (e.g. Ahn \& Shapiro 2007; Whalen et al. 2007).  Thus, it appears likely that intermittent local radiative feedback effects may not be decisive in suppressing the formation of Pop~III stars during the assembly of the first galaxies (see also Wise \& Abel 2007a). 

The properties of the first galaxies are, however, more likely related to the global LW background, as it has been shown that a strong, persistent LW background can suppress star formation in the minihaloes which eventually are merged to form the first galaxies (e.g. Dekel \& Rees 1987; Haiman et al. 1997; Bromm \& Larson 2004).  If star formation in these small systems can be suppressed effectively by the LW background, the first protogalaxies are much more likely to be kept metal-free.  However, if star formation is not easily suppressed in minihaloes, either because the LW background is weak or if star formation can continue unimpeded despite strong LW feedback, then the first galaxies will likely form from already metal-enriched gas. Many studies have sought to estimate the global LW background during the epoch of the first stars (Haiman et al. 1997, 2000; Ricotti et al. 2002a,b; Glover \& Brand 2003; Yoshida et al. 2003; Wise \& Abel 2005; Greif \& Bromm 2006). Additionally, the impact of a constant LW background on Pop III star formation has been investigated, suggesting that, while global LW feedback does not completely suppress star formation in minihaloes, it can delay such star formation (Machacek et al. 2001; Mesinger et al. 2006; O'Shea \& Norman 2008; Wise \& Abel 2007b). 

In this paper, we investigate the impact of the build-up of a LW background on star formation in systems which later evolve into the first galaxies at redshift $z$ $\ga$ 10.  In Section 2, we develop analytical models which self-consistently couple the evolution of the LW background to that of the Pop III star formation rate (SFR) at $z$ $\ga$ 15.  We describe our implementation of the LW background in cosmological simulations in Section 3.  We report the outcome of simulations designed to test the hypothesis of a self-regulated LW background in Section 4, while in Section 5, we simulate the assembly of the first galaxy under the influence of such a self-regulated LW background. Finally, in Section 6 we discuss the implications of our findings for the galaxy formation process, and we conclude in Section 7.

\section {The Lyman-Werner Background}

The LW background radiation field is produced by sources, either stars or miniquasars, at cosmological distances, beginning with the formation of the first stars at the end of the cosmic dark ages.  This radiation field acts to delay star formation by dissociating H$_2$ molecules, which are the primary coolants allowing primordial gas to collapse and form stars in minihaloes.  Thus, to fully address both the production of the LW background radiation and the Pop III star formation which is sensitively coupled to it would require a cosmological simulation resolving the collapse of gas into minihaloes over enormous volumes of the Universe. As carrying out simulations over such large cosmological scales, while simultaneously resolving the collapse of gas in minihaloes, is still prohibitively expensive, we revert to simple analytical estimates for the self-consistent build-up of the LW background which draw on the results of detailed numerical simulations of Pop III star formation and radiative feedback.  The simple self-consistent analytical model that we develop in this Section may serve as a first order approximation to a more detailed self-consistent model for the build-up of the LW background and of the Pop III star formation rate in minihaloes at redshifts $z$ $\ga$ 15.

\subsection{The Critical LW Flux}
We first estimate the LW flux required to significantly delay the formation of Pop III stars in minihaloes.  We note that for a typical case, following the extensive results of detailed simulations of Pop III star formation (see e.g. Bromm et al. 2002; Yoshida et al. 2003), the primordial gas first collapses adiabatically into a minihalo until it reaches a temperature $T$ $\ga$ 10$^3$ K and density $n$ $\sim$ 1 cm$^{-3}$.  As the evolution of the gas is adiabatic up to this point, the cooling properties of the gas play a role in its evolution only at higher densities.  In particular, for the gas to continue collapsing on the way to forming a Pop III star, in accordance with the results of detailed simulations, a sufficient fraction of the coolant H$_2$ is required (Tegmark et al. 1997).  In the presence of a LW radiation field, however, H$_2$ may be destroyed at a rate higher than it can be produced, effectively delaying the collapse of the gas.  

We may estimate the flux of the radiation field required to achieve this delay by comparing the formation time of H$_2$ molecules, at $T$ $\sim$ 10$^3$ K and $n$ $\sim$ 1 cm$^{-3}$, to the dissociation time of H$_2$ molecules as determined by the LW flux.  For the dissociation time, we have (Abel et al. 1997)

\begin{equation}
t_{\rm diss} \sim 3 \times 10^4 J_{\rm LW }^{-1} {\rm \ yr} \mbox{\ ,}
\end{equation} 
where $J_{\rm LW}$ is in units of 10$^{-21}$ erg s$^{-1}$ cm$^{-2}$ Hz$^{-1}$ sr$^{-1}$. Equating to the formation time found in our simulations, $t_{\rm form}$ $\sim$ 8 $\times$ 10$^5$ yr, yields $t_{\rm diss}$ = $t_{\rm form}$ when the LW flux is $J_{\rm LW,crit}$ $\sim$ 0.04, which we define as the critical flux necessary to significantly delay Pop III star formation in minihaloes, roughly consistent with numerous simulation results (see Machacek et al. 2001; Yoshida et al. 2003; Mesinger et al. 2006; O'Shea \& Norman 2008; Wise \& Abel 2007b).

\begin{figure}
\vspace{2pt}
\epsfig{file=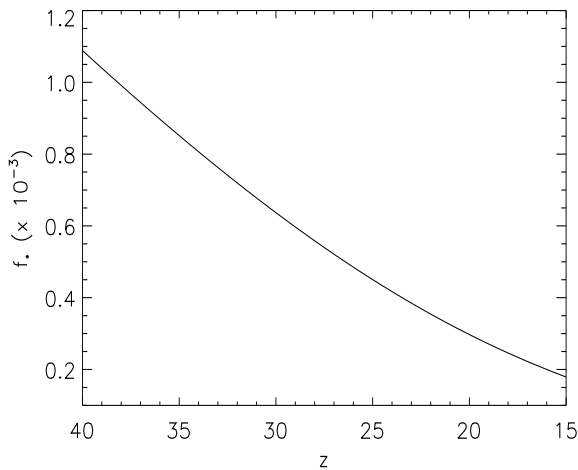,width=8.5cm,height=7.0cm}
%VB: small change here!
\caption{The average star formation efficiency $f_*$ for Pop~III star formation in minihaloes, assuming that a single 100 M$_{\odot}$ star forms in each collapsed minihalo with a virial temperature 2 $\times$ 10$^3$ K $<$ $T_{\rm vir}$ $<$ 10$^4$ K. Here $f_*$ is defined as the fraction of collapsed baryons converted into Pop~III stars. The average is carried out over the Sheth-Tormen mass function.}
\end{figure}

\subsection{The Maximum Background Flux}
%VB: I ve added a statement that f_star is an average over the ST mass function.
%    Please check whether you agree!
As a first step in developing our estimate of a self-consistent Pop III star formation rate and LW background flux, we calculate the maximum value for $J_{\rm LW}$, as a function of redshift $z$, that we expect to be produced by Pop III stars forming in minihaloes at $z$ $\ga$ 15.  We use the Sheth-Tormen formalism (see Sheth et al. 2001), with the cosmological parameters derived from the {\it Wilkinson Microwave Anisotropy Probe} ({\it WMAP}) third year data (Spergel et al. 2007), to find the mass fraction collapsed in minihaloes at these high redshifts. In comparison, the collapsed fraction in more massive haloes is very small, allowing us to neglect star formation that may be taking place in systems with virial temperatures $T_{\rm vir}$ $\ga$ 10$^4$ K (see Section 2.3). Most studies of primordial star formation in minihaloes find that single stars with masses of the order of 100 M$_{\odot}$ form in these systems (Abel et al. 2002; Bromm \& Loeb 2004; Yoshida et al. 2006; Gao et al. 2007). To estimate the maximum star formation rate we assume that each minihalo with a virial temperature $T_{\rm vir}$ $\ga$ 2 $\times$ 10$^3$ K (see e.g. Yoshida et al. 2003) hosts a 100 M$_{\odot}$ Pop III star which lives for $t_{\rm *}$ = 3 Myr (e.g. Schaerer 2002).  The comoving star formation rate of Pop III stars in minihaloes, in the absence of the negative feedback due to LW radiation, can then be estimated to be

\begin{eqnarray}
{\rmn SFR}_{\rmn III, max} & \sim & 10^{-3}\left(\frac{f_*}{10^{-3}}\right) \left(\frac{\left|dF/dz\right|}{2 \times 10^{-3}}\right) \nonumber \\ 
                  & \times & \left(\frac{1+z}{16}\right)^{\frac{5}{2}}{\rmn M_{\odot} yr^{-1} Mpc^{-3}} \mbox{\ ,}
\end{eqnarray}
where $f_*$ is the fraction of collapsed baryons converted into Pop~III stars, averaged over the Sheth-Tormen mass function, and $\left|dF/dz\right|$ is the change in the collapse fraction in minihaloes with redshift derived from the Sheth-Tormen formalism.  We calculate $f_*$ assuming that every minihalo with 2 $\times$ 10$^3$ K $<$ $T_{\rm vir}$ $<$ 10$^4$ K hosts a single Pop~III star with a mass of 100 M$_{\odot}$, and the resulting evolution of $f_*$ with redshift is shown in Fig. 1.  Due to the increase of the average mass of such minihaloes with decreasing redshift, the fraction of collapsed baryons which goes into stars drops with redshift.
For the following argument, it is convenient to compute a physical (proper) number density of Pop~III stars in minihaloes as a function of redshift $z$, which we find to be 
\begin{equation}
n_{\rmn III, max}\sim 6 \times 10^4 \left(\frac{f_*}{10^{-3}}\right)\left(\frac{\left|dF/dz\right|}{2 \times 10^{-3}}\right)\left(\frac{1+z}{16}\right)^{\frac{11}{2}}{\rmn Mpc^{-3}} \mbox{\ ,}
\end{equation} 
assuming a constant Pop~III mass of 100 M$_{\odot}$.

%VB: I merged the paragraph above with the preceding one; it s too small otherwise.
%VB: small changes in next sentence!
To calculate the LW background generated in the limiting case that {\it every} minihalo hosts Pop~III star formation, we next estimate the maximum physical distance traveled by a LW photon before being absorbed by a Lyman series transition in atomic hydrogen, which will be abundant at these epochs, when the Universe has not yet been substantially reionized (e.g. Haiman et al. 1997; Mackey et al. 2003). This distance corresponds to that of light traveling for a time needed to redshift the Ly$\beta$ line of neutral hydrogen, at 12.1~eV, to the low-energy end of the LW band at 11.2~eV. At this point essentially all LW photons, with energies between 11.2 and 13.6 eV, will have been absorbed by neutral hydrogen.  The distance to the LW horizon is thus 

\begin{equation}
{r_{\rm max}} \sim 10 \left(\frac{1 + z}{16}\right)^{-\frac{3}{2}} {\rm Mpc} \mbox{\ ,}
\end{equation}  
again given in physical units.  We note that this distance is an upper limit to how far a LW photon can propagate; for a more detailed calculation of the frequency-dependent propagation of LW radiation see Haiman et al. (2000).  Using the number density of Pop~III stars in equation (3), we integrate the flux from all the stars within a sphere of radius $r_{\rm max}$, obtaining the following estimate for the maximum background LW flux: 

\begin{equation}
{J_{\rmn LW, max}} \sim 2 \left(\frac{f_*}{10^{-3}}\right) \left(\frac{\left|dF/dz\right|}{2 \times 10^{-3}}\right)\left(\frac{1 + z}{16}\right)^4 \mbox{\ .}
\end{equation} 
Following Johnson et al. (2007), we here assume that the LW photons are emitted with a blackbody spectrum at 10$^5$~K from a 100 M$_{\odot}$ Pop~III star, as modeled by Bromm et al. (2001b). In Fig.~2, we show the redshift evolution of $J_{\rmn LW, max}$, along with the corresponding SFR$_{\rmn III, max}$, assuming that $f_*$ evolves as shown in Fig.~1.

\begin{figure}
\vspace{2pt}
\epsfig{file=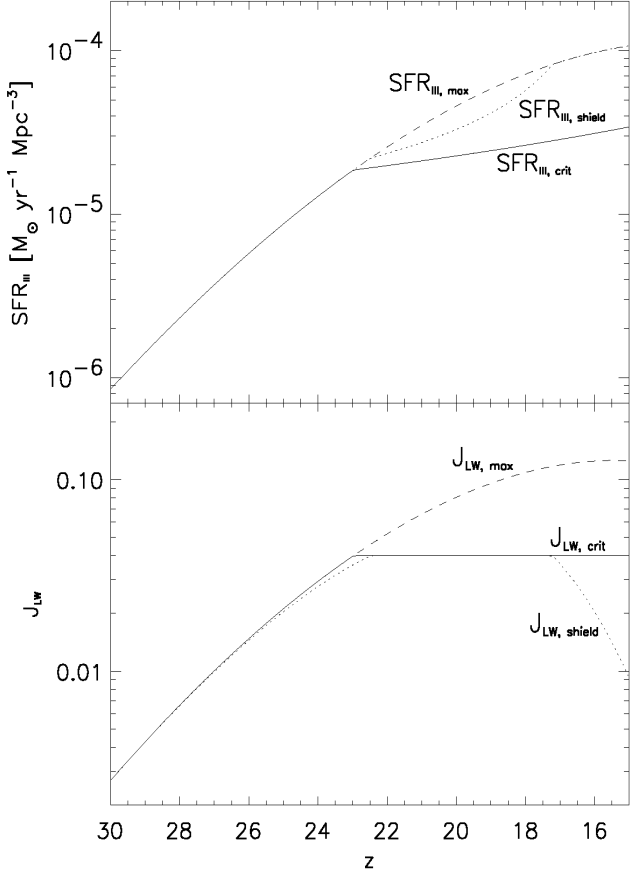,width=8.5cm,height=12.0cm}
%VB: I ve made all SFRs roman-style (\rm), in accordance with custom.
\caption{The Pop~III star formation rates ({\it top panel}) and the corresponding LW background fluxes ({\it bottom panel}) for the three cases discussed in Sections 2.2 and 2.3.  The maximum possible LW background, $J_{\rm LW, max}$, is generated for the case that every minihalo with a virial temperature $T_{\rm vir}$ $\geq$ 2 $\times$ 10$^3$ K hosts a Pop~III star, without the LW background in turn diminishing the SFR (dashed lines). The self-regulated model considers the coupling between the star formation rate, ${\rm SFR}_{\rm III, crit}$, and the LW background built-up by it, $J_{\rm LW, crit}$ (solid lines). The minimum value for the LW background, $J_{\rm LW, shield}$, is produced for the case of a high opacity through the relic H~II regions left by the first stars and a corresponding star formation rate ${\rm SFR}_{\rm III, shield}$, higher than ${\rm SFR}_{\rm III, crit}$ and approaching ${\rm SFR}_{\rm III, max}$ (dotted lines). 
We expect that the true Pop~III SFR in minihaloes has a value intermediate between ${\rm SFR}_{\rm III, crit}$ and ${\rm SFR}_{\rm III, shield}$.}
\end{figure}

We note that this idealized approach somewhat overestimates the LW background from 100 M$_{\odot}$ primordial stars, as many of the LW photons will be absorbed over shorter distances by the higher order Lyman series transitions corresponding to higher LW photon energies (e.g. Haiman et al. 2000).  Nevertheless, we expect that the above calculation gives a useful estimate for the maximum $J_{\rm LW}$ generated by Pop~III star formation in minihaloes.

\subsection{Self-consistent Models of the LW Background}
In our calculation of the maximum LW background, we assumed that the Pop~III SFR was independent of the level of the LW background generated.  However, as shown in Fig.~2, the value of $J_{\rmn LW, max}$ can exceed that necessary to significantly delay star formation (e.g. Machacek 2001; Mesinger 2006; O'Shea \& Norman 2008; Wise \& Abel 2007b).  Therefore, the model presented in Section 2.2 is inconsistent in that the SFR required to produce the maximum LW background cannot be maintained under such a strong background. Therefore, we now modify our model to more accurately couple the SFR to the LW background.

\subsubsection{Self-regulation}
%VB: small changes to this subsection!
The sources that produce the LW flux at redshifts $z$ $\ga$ 15 may be stars or miniquasars, in principle. However, the accretion onto massive black holes resulting from the collapse of Pop~III stars is not efficient in general (Johnson \& Bromm 2007; Pelupessy et al. 2007), although it may be in relatively rare cases (see Li et al. 2007), and so the first miniquasars were likely not an important source of LW radiation in the early Universe. We further assume that the dominant contribution to the LW background at $z\ga 15$ is due to Pop~III star formation in minihaloes, and we here neglect atomic cooling haloes as additional sources. At these redshifts, $\ga$ 90 percent of the collapsed mass is in minihaloes which, if they host star formation, are expected to host single massive Pop III stars (Abel et al. 2002; Yoshida et al. 2006; Gao et al. 2007). Therefore, if stars form in atomic-cooling haloes (e.g. Oh \& Haiman 2002) with the same efficiency $f_*$ with which stars form in minihaloes, the vast majority of stars will form in minihaloes at $z$ $\ga$ 15, thus justifying our assumption.  

However, the star formation efficiency $f_*$ in atomic-cooling haloes is not well-constrained at $z$ $\ga$ 15. Recently, Ricotti et al. (2008) have argued that $f_*$ can take a range of values, from $\sim$ 10$^{-3}$ to $\sim$ 10$^{-1}$,
roughly scaling as the square of the halo mass.
Because most of the collapsed mass that goes into atomic-cooling haloes will reside in the least massive systems, we take this as evidence that the 
mass-weighted average of $f_*$ is likely near the lower end of this range, which is similar to minihalo efficiencies. In Fig. 3 we show the fraction of LW photons, $f_{\rm LW, mini}$, emitted 
by Pop III stars formed in minihaloes with a star formation rate given by ${\rm SFR}_{\rm III,crit}$, for various assumptions about the efficiency of star formation and the IMF of stars formed 
in atomic-cooling haloes.  While for minihalo star formation we assume that single 100 M$_{\odot}$ stars form with an efficiency as given in Fig. 1, we consider values of $f_*$ = 10$^{-1}$, 10$^{-2}$, and 10$^{-3}$ for atomic cooling haloes. Furthermore, we consider two IMFs for atomic-cooling haloes: a standard Salpeter IMF for the case of an already metal-enriched (Pop~II) system, and a Pop~III.2 IMF postulating that metal-free stars form with masses of the order of 10 M$_{\odot}$ (Johnson \& Bromm 2006; see McKee \& Tan 2007 for the terminology). The latter case is the predicted star formation mode in atomic cooling haloes, provided that they remain pristine (e.g. Johnson \& Bromm 2006; Greif et al. 2008). For the Salpeter IMF, the number of LW photons emitted per baryon in stars is taken to be $\eta_{\rm LW}$ = 4 $\times$ 10$^{3}$, and for the Pop~III.2 case, we assume $\eta_{\rm LW}$ = 2 $\times$ 10$^{4}$ (see Greif \& Bromm 2006). Because our model of self-regulated star formation assumes that minihaloes produce the dominant portion of the LW background, it will only be valid for redshifts where $f_{\rm LW, mini}$ $\ga$ 0.8. Assuming $f_*$ $\ga$ 10$^{-3}$ in atomic-cooling haloes, we thus expect that our self-regulated model of star formation in minihaloes is valid at redshifts $z$ $\ga$ 15 for a Salpeter IMF, and at redshifts $z$ $\ga$ 18 for a Pop~III.2 IMF. As suggested by the results presented in Section 5, the IMF in such systems may transition from a Pop~III IMF to a Pop~II IMF after the formation of only one or a few massive Pop~III stars.

With the assumption that Pop~III stars in minihaloes contribute the dominant portion of the LW background, the production of the LW background becomes a purely self-regulated process:  any increase in the star formation rate will translate into an increase in the LW background, which will in turn suppress the star formation rate, and vice versa.  
While a prohibitively expensive simulation would be required to self-consistently track the build-up of the LW background in this way (but see Yoshida et al. 2003), we suggest that a self-consistent flux would be similar to the value that we found for the critical LW flux, $J_{\rm LW, crit}$.  As explained in Section 2.1, a LW flux much below this value will not efficiently delay the formation of molecules in minihaloes and so will only slightly suppress the star formation rate; in turn, a LW flux much above this value will delay the star formation rate substantially (e.g. O'Shea \& Norman 2008; Wise \& Abel 2007b).  

The star formation rate expected under the assumption that the LW background must not exceed $J_{\rmn LW,crit}$ can be derived from the Sheth-Tormen formalism.  We follow our treatment for the maximum LW flux from Pop~III star formation in minihaloes (Section 2.2), now requiring that $J_{\rm LW, crit} = 0.04$ is not exceeded, as shown in Fig.~2. We then find the fraction $f_{\rm reg}$ of minihaloes that must form stars in order to produce this self-regulated LW flux from the simple relation

\begin{equation}
f_{\rm reg}(z) \sim \frac{J_{\rm LW, crit}(z)}{J_{\rm LW, max}(z)} \mbox{\ .}  
\end{equation}
Assuming that Pop~III stars form in minihaloes, if not prevented from doing so by negative feedback, as single, massive stars (e.g. O'Shea \& Norman 2008; Wise \& Abel 2007b), it is the number of minihaloes at a given redshift which determines the SFR. We thus infer that this critical rate, SFR$_{\rm III, crit}$, is suppressed, at most, by a fraction $f_{\rm reg}$ compared to the maximum possible Pop III rate in minihaloes, SFR$_{\rm III, max}$. Because our value for $J_{\rm LW, max}$ is an upper limit (see Section 2.2), the corresponding suppression factor, $f_{\rm reg}$, is a lower limit.

The calculation presented here gives a simple estimate of the Pop III SFR at $z$ $\ga$ 15, accounting for the coupling of the SFR to the LW background generated by Pop~III stars.  In Section~4, we report on a simulation carried out to test the validity of this model as a first approximation to a more detailed self-consistent model of the Pop III star formation rate history. 

\begin{figure}
\vspace{2pt}
\epsfig{file=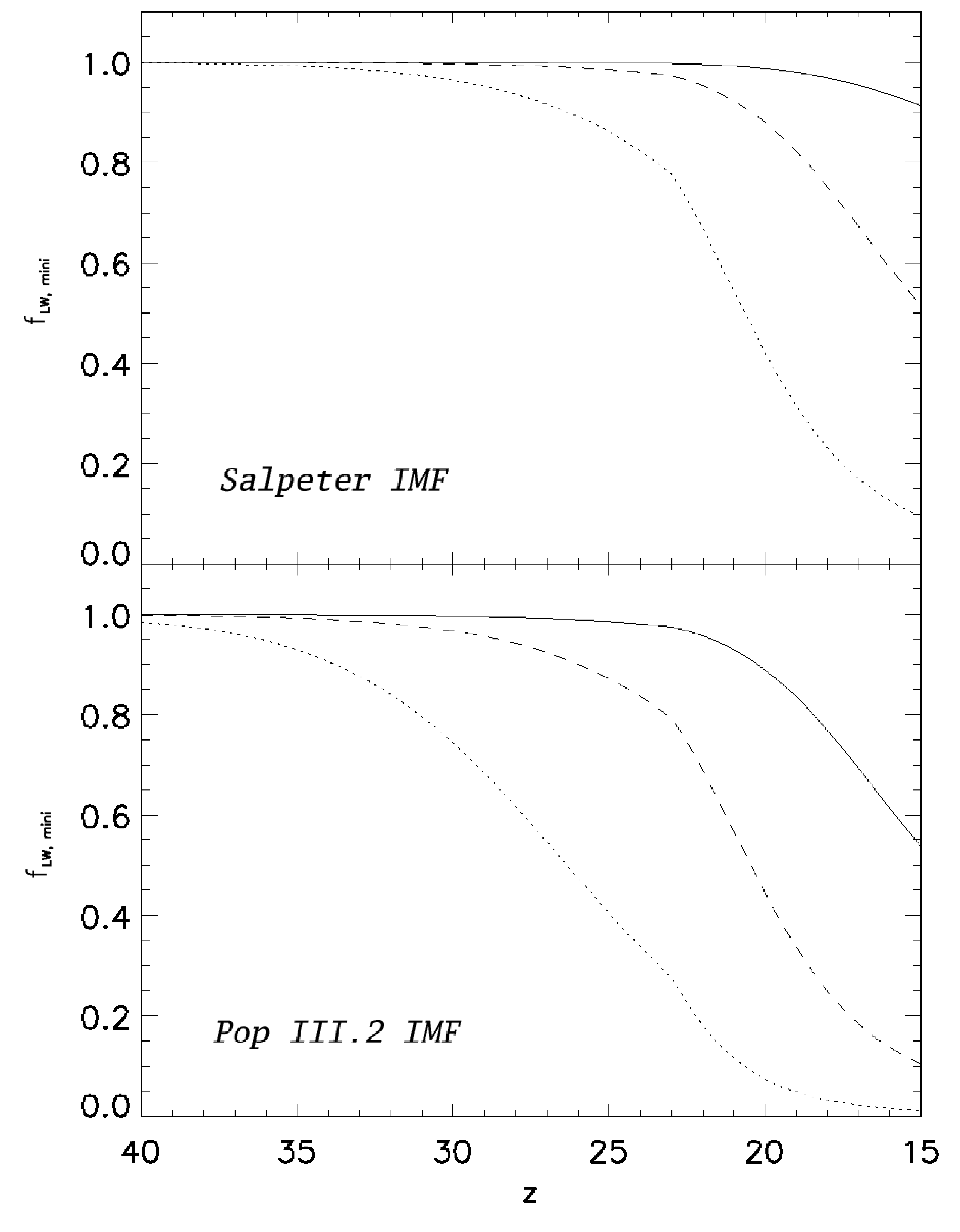,width=8.5cm,height=12.0cm}
%VB: small changes to caption!
\caption{
The fraction of the total LW flux contributed by Pop III stars formed in minihaloes $f_{\rmn LW, mini}$ in our self-regulated model (see Section 2.3.1), for various choices of the star formation efficiency $f_*$ in atomic-cooling haloes, which we assume are not subject to LW feedback.  Atomic-cooling haloes in which Pop II stars form with a Salpeter initial mass function (IMF) ({\it top panel}) produce a smaller fraction of LW photons, for a given star formation efficiency, than do atomic-cooling haloes which host Pop~III.2 stars with masses of the order of 10 M$_{\odot}$ ({\it bottom panel}).  The solid lines correspond to a star formation efficiency of $f_*$ = 10$^{-3}$ in atomic-cooling haloes, while the dashed and dotted lines correspond to $f_*$ = 10$^{-2}$ and 0.1, respectively.  
Our self-regulated model assumes that Pop~III stars formed in minihaloes emit the dominant portion of LW photons ($f_*$ $\geq$ 0.8).  
For different choices of the IMF and the star formation efficiency, our model will be valid down to different redshifts.
}
\end{figure}

\subsubsection {Shielding from Relic H~II Regions}
%VB: some changes to this subsection!
While massive Pop III stars formed in minihaloes create large H~II regions, with (proper) radii of a few kpc (Alvarez et al. 2006; Abel et al. 2006), these stars live for only $\la$ 3 Myr, at which point the relic H~II regions that are left behind begin to cool and recombine. The conditions in such relic H~II regions are conducive to the rapid formation of H$_2$ and HD molecules (Nagakura \& Omukai 2005; O'Shea et al. 2005; Johnson \& Bromm 2006; Yoshida et al. 2007). Johnson et al. (2007) showed that the resulting opacity to LW photons, $\tau_{\rm LW}$, can be significant, exceeding unity across just a single relic H~II region.  Furthermore, these authors found that the high H$_2$ fraction formed in these regions generally remains high, as molecules quickly reform in between the intermittent episodes of exposure to LW radiation produced by nearby, short-lived Pop III stars. 

To evaluate the effect of the opacity through relic H~II regions on the build-up of the LW background, we now repeat the calculation in Section 2.2, this time including a maximal optical depth to LW photons through relic H~II regions, $\tau_{\rm LW, max}$.  We employ an iterative procedure 
in order to determine self-consistently, at each redshift, the star formation rate, the optical depth to LW photons, and the LW background flux, all of which are dependent on each other.  To begin, we initialize the calculation at $z$ = 50, at which time we assume that $\tau_{\rm LW, max}$ = 0, the star formation rate is equal to the maximal rate ${\rm SFR}_{\rm III, max}$, and the LW background flux is equal to the maximal value $J_{\rm LW, max}$, and we integrate forward in time.   

To calculate the cosmological average $\tau_{\rm LW, max}$ we adopt the results of Haiman et al. (2000), who find that at $z$ = 15 the average 
LW optical depth is $\tau_{\rm LW}$ $\sim$ 0.04 for a uniform H$_2$ fraction of $f_{\rm H_2}$ = 2 $\times$ 10$^{-6}$.  Consistent with the results of Haiman et al., this cosmological
average optical depth scales as 

\begin{equation}
\tau_{\rm LW}  \propto r_{\rm max} n_{\rm H_2} \propto \left(1 + z \right)^{\frac{3}{2}} \mbox{\,}        
\end{equation}
where $n_{\rm H_2}$ is the average number density of H$_2$ in the intergalactic medium (IGM).  

As Pop~III star formation in minihaloes proceeds, the volume-filling fraction of relic H~II regions increases 
with the number of Pop~III stars which form and rapidly die. Johnson \& Bromm (2007) found that the high residual free-electron
 fraction in relic H~II regions can persist for of the order of 500~Myr, in turn driving the vigorous production of H$_2$ molecules. These regions are thus expected to maintain 
a high H$_2$ fraction for over a Hubble time, and their abundance is therefore assumed to be proportional to the cumulative star formation history and to only increase with time. 
We calculate the volume-filling fraction of relic H~II regions, $f_{\rm HII}$, by assuming that each Pop~III star which forms in a minihalo leaves behind a relic H~II region.  We find

\begin{equation}
{f_{\rm HII}} \sim \frac{4 \pi}{3} \left[4 {\rm kpc} \left(\frac{1 + z}{20}\right)^{-1}\right]^3 n_{\rm HII} \mbox{\ ,}
\end{equation} 
where we have assumed that each Pop III star leaves behind a relic H~II region with a radius of $\sim$ 4 kpc $[(1+z)/20]^{-1}$ in which the recombination time is longer than the lifetime of the star (see Abel et al. 2006; Alvarez et al. 2006; Johnson et al. 2007; Yoshida et al. 2007). The (physical) number density of relic H~II regions, $n_{\rm HII}$, taken to be proportional to the collapse fraction in minihaloes as explained above, is given by

\begin{eqnarray}
{n_{\rm HII}} \sim 10^{6}  f_{\rm shield} \left(\frac{f_*}{10^{-3}}\right) \left(\frac{F(z)}{6 \times 10^{-3}}\right)\left(\frac{1 + z}{16}\right)^3  {\rm Mpc}^{-3} \mbox{\ .}
\end{eqnarray} 
Here $F(z)$ is the fraction of collapsed mass in minihaloes as a function of redshift, given by the Sheth-Tormen formalism, with the fiducial value of $F(z)$ $\sim$ 6 $\times$ 10$^{-3}$ at $z$ $\sim$ 15, while $F(z)$ $\sim$ 10$^{-3}$ at $z$ $\sim$ 20. 
We account for the suppression in the SFR due to a shielded LW background by including the factor $f_{\rm shield}$, defined as the ratio of the integrated SFR in the self-regulated model including an IGM optical depth, ${\rm SFR}_{\rm III, shield}$, to be estimated below, to the integrated maximum SFR, ${\rm SFR}_{\rm III, max}$, as described in Section 2.1:

\begin{equation}
f_{\rmn shield} = \frac{\int_z^{\infty} {\rmn SFR}_{\rmn III, shield}(z')dz'}{\int_z^{\infty} {\rmn SFR}_{\rmn III, max}(z')dz'} \mbox{\ .} 
\end{equation}    
We can then re-write the volume-filling factor of H~II regions as

\begin{equation}
f_{\rm HII} \sim 0.1  f_{\rm shield} \left(\frac{f_*}{5 \times 10^{-4}}\right)\left(\frac{F(z)}{6 \times 10^{-3}}\right) \mbox{\ .}
\end{equation} 
For simplicity, we use a constant average value of $f_*$ = 5 $\times$ 10$^{-4}$ in equation (11), consistent with the values shown in Fig. 1. 

To obtain the maximum average IGM optical depth to LW photons, we follow the results of Haiman et al. (2000) along with equation (7), this time accounting for an elevated H$_2$ abundance within a fraction $f_{\rm HII}$ of the volume of the Universe.  We thus find

\begin{equation}
\tau_{\rm LW, max} = 4 \left(\frac{f_{\rm HII}}{0.5}\right) \left(\frac{f_{\rm H_2}}{10^{-4}}\right) \left(\frac{1 + z}{16} \right)^{\frac{3}{2}} \left(\frac{\delta}{10}\right)\mbox{\ ,}
\end{equation}
where we use $f_{\rm H_2}$ = 10$^{-4}$ as an upper limit for the average H$_2$ fraction in relic H~II regions and an average overdensity in these regions of $\delta = 10$, as estimated from the simulation presented in Johnson et al. (2007).  

Next, we self-consistently calculate the minihalo Pop~III SFR, again assuming that the LW background flux must not exceed $J_{\rm LW, crit}$, but this time taking into account the optical depth $\tau_{\rm LW, max}$ through relic H~II regions.  The fraction of minihaloes in which Pop~III stars are able to form is then given by    

\begin{equation}
f_{\rm \tau}(z) = {\rm min} \left( 1, \frac{J_{\rm LW, crit}(z)}{J_{\rm LW, max}(z)} e^{\tau_{\rm LW, max}} \right) \mbox{\ ,}
\end{equation}
analogous to the no-shielding case in equation (6). The intuition here is that to compensate for the loss due to the IGM opacity, a larger SFR is required to maintain a LW background at the critical level, compared to ${\rm SFR}_{\rm III, max}$; however, the SFR cannot exceed the maximum one. We thus find for the SFR of Pop~III stars in minihaloes, including the effects of both a LW background and the IGM optical depth to LW photons

\begin{equation}
{\rm SFR}_{\rm III, shield} = f_{\rm \tau}(z) {\rm SFR}_{\rm III, max} \mbox{\ .} 
\end{equation}  
Finally, the corresponding LW background flux is

\begin{equation}
J_{\rm LW, shield} = J_{\rm LW, crit} \times {\rm min}\left(1, \frac{J_{\rm LW, max}(z)}{J_{\rm LW, crit}(z)} e^{-\tau_{\rm LW, max}}\right)\mbox{\ ,}
\end{equation}
as plotted in Fig. 2. The optical depth $\tau_{\rm LW, max}$ produced in this model is shown in Fig. 4.  We interpret $J_{\rm LW, shield}$ to be the minimal value for the LW background flux at $z$ $\ga$ 15.  Correspondingly, we find that ${\rm SFR}_{\rm III, shield}$ is an upper limit to the Pop~III SFR in minihaloes at these redshifts.

\begin{figure}
\vspace{2pt}
\epsfig{file=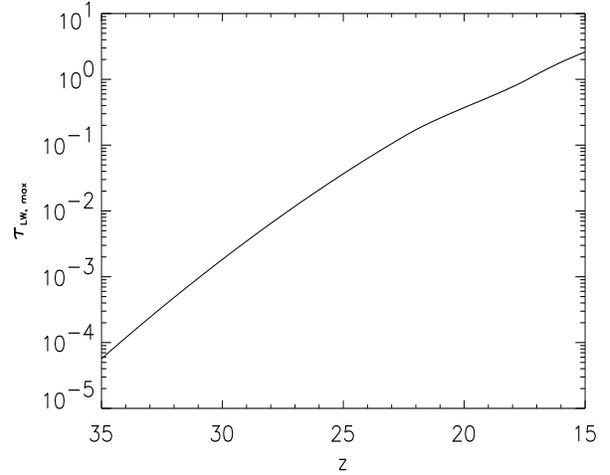,width=8.5cm,height=7.cm}
%VB: small changes to caption here!
\caption{The maximum IGM optical depth to LW photons, $\tau_{\rm LW, max}$, 
due to the high fraction of H$_2$ in the relic H~II regions. This corresponds to ${\rm SFR}_{\rm III, shield}$ and a LW background flux of $J_{\rm LW, shield}$.  Note that locally the optical depth can be greater than this average, exceeding unity through a single H~II region, owing to self-shielding of H$_2$ (see Johnson et al. 2007).
}
\end{figure}

 In reality, the clustering of Pop~III sources, coupled with the increasing SFR towards lower redshifts, will likely lead to the efficient photodissociation of H$_2$ molecules in some relic H~II regions even at $z \ga 15$, as well as to overlap between relic H~II regions as $f_{\rmn HII}$ grows with time.  Both effects will lower the opacity to LW photons through the IGM, and the global average LW background will eventually exceed the level $J_{\rm LW}$ $\sim$ 10$^{-3}$ needed for the dissociation of H$_2$ molecules 
in these regions to outpace their formation (Johnson \& Bromm 2007), although we note that we have not accounted for the local optical depth to LW photons 
through individual relic H~II regions, which can exceed unity through a single relic H~II region (Johnson et al. 2007).  Finally, we note that the emission of X-rays from miniquasars may, in principle, lead to a high opacity to LW photons, as it has been shown that the partial ionization from X-rays can enhance the generation of H$_2$ in the IGM (e.g. Kuhlen \& Madau 2005).  However, as noted above, it is not clear that the accretion rates onto Pop~III relic black holes are sufficiently high for this mechanism to dominate the generation of LW opacity (e.g. Pelupessy et al. 2007).

\subsection{The Expected Pop III Star Formation Rate}
We expect the actual LW background at $z$ $\ga$ 15 to lie between $J_{\rm LW, shield}$ and $J_{\rm LW, crit}$. 
The corresponding Pop III SFR in minihaloes which generates this LW background likely falls between SFR$_{\rm III, crit}$ and SFR$_{\rm III, shield}$, 
since the shielding provided by relic H~II regions can keep $J_{\rm LW} \la J_{\rm LW, crit}$. In turn, this will allow to keep SFR$_{\rm III}$ $\ga$ SFR$_{\rm III, crit}$, owing to the diminished strength of the negative LW feedback.  
In the remainder of the present work we will take $J_{\rm LW, crit}$ as a plausible maximum value for the LW background flux, and we will use this value as an input in our simulations of galaxy formation, thereby obtaining lower limits to the amount of star formation that may take place in the assembly of these systems at $z$ $\ga$ 15.  We note that, although we follow the assembly of a galaxy to $z$ $\ga$ 12, the star formation that does take place in our simulation occurs at $z$ $\ga$ 15. Thus, our simple assumption here that $J_{\rm LW}$ = $J_{\rm LW, crit}$ even at $z$ $\la$ 15 does not impact the results of our simulations significantly.

\section{Methodology of Simulations}
With the goal of estimating the degree to which Pop III star formation in minihaloes affects the properties of the gas from which the first galaxies form at $z$ $\ga$ 10, we have carried out two cosmological hydrodynamics simulations which track the detailed evolution of the primordial gas under the influence of a LW background radiation field. The first simulation is designed to test the validity of our self-regulated model for the build-up of the LW background, and is described further in Section 4.  The second simulation follows the assembly of a galaxy virializing at $z$ $\ga$ 10, assuming a critical LW background, $J_{\rm LW, crit}$. This simulation is described in Section 5.     

As with previous work, for our three-dimensional numerical simulations we employ the parallel version of GADGET (version 1), which includes a tree (hierarchical) gravity solver combined with the smoothed particle hydrodynamics (SPH) method for tracking the evolution of gas (Springel et al. 2001; Springel \& Hernquist 2002).  Along with H$_2$, H$_2^{+}$, H, H$^-$, H$^+$, e$^-$, He, He$^{+}$, and He$^{++}$, we have included the five deuterium species D, D$^+$, D$^-$, HD and HD$^+$, using the same chemical network as in Johnson \& Bromm (2006, 2007).  The specific initial conditions for our simulations are described separately in Sections 4 and 5.

\subsection{Sink Particles}
The primary result obtained from the present simulations is the number of Pop~III stars which are able to form in minihaloes at $z$ $\ga$ 15 during the assembly of a galaxy virializing at $z$ $\ga$ 10. To track the number and location of the formation sites for Pop III stars, we allow sink particles to form when the gas has collapsed to high densities. Specifically, sinks are created when the gas has collapsed to a threshold density $n_{\rm res}$, above which the mass resolution of our simulation is insufficient to reliably follow the continued 
collapse of the gas, as explained further in Johnson et al. (2007).  We note that the time at which a sink particle forms will precede the time at which a star is able to form by roughly a free-fall time in gas with a density $n_{\rm res}$; while this can introduce an error of $\la$ 10 Myr in the precise timing of star formation, the conclusions that we derive from the results presented here are not significantly impacted. Once having formed, the sink particles act as tracers of star formation sites, each sink representing the formation site of a single Pop~III star. We focus our attention on when and where these sinks emerge in order to estimate the number of stars formed, and to gauge the feedback effects from these stars. 

\subsection{Implementation of Background Radiation}
Motivated by the considerations presented in Section 2 and further justified in Section 4, we have included in our simulations the self-regulated model of the LW background radiation field, in the form of a global H$_2$ dissociation rate.  The LW background is taken to be a function of redshift, its evolution given by $J_{\rm LW, crit}(z)$, as shown in Fig.~2. The dissociation rate is

\begin{equation}
k_{\rm diss}(z) = 10^{-12} J_{\rm LW, crit}(z) {\rm s}^{-1}\mbox{\ .}
\end{equation}

In addition to the photodissociation of H$_2$, the photodestruction of other species can, in principle, affect the cooling properties of the primordial gas.  The main reaction producing H$_2$ being H$^-$ + H $\to$ H$_2$ + e$^-$ (e.g. Peebles \& Dicke 1968; Lepp \& Shull 1984), the photodestruction of H$^-$ is a process that might lead to suppression of the H$_2$ fraction of the gas (Chuzhoy et al. 2007; see also Glover et al. 2006).  However, we find the photodissociation of H$^-$ to be unimportant for the suppression of H$_2$ in collapsing minihaloes.  As discussed in Section 2.1, the gas in such minihaloes will adiabatically collapse to densities of $\ga$ 1 cm$^{-3}$ and temperatures of $\ga$ 10$^3$ K, even if the fractions of H$_2$ or H$^-$ are low.  Following this adiabatic collapse, the fate of the gas does depend sensitively on the abundances of these species, and it is only at this stage at which the photodissociation of H$^-$ might begin to have an effect.  We find that the chemical species at this stage, absent strong shocks or photoionization, are roughly in equilibrium when we include the effect of the LW background, and therefore suppression of the H$_2$ abundance can be directly tied to the abundance of H$^-$ at this stage.  At this temperature and density, we find that the H$^{-}$ formation time is of the order of years.  We find that the minimal photodissociation timescale of H$^-$, produced for the case of the maximal LW background $J_{\rm LW, max}$ (see Section 2.2), is of the order of $\ga$ 10$^2$ yr, roughly two orders of magnitude higher than the formation timescale.  Thus, the H$^-$ fraction will not be significantly affected by a background radiation field, and in turn the abundance of H$_2$ will not be significantly affected, allowing us to neglect the photodissociation of H$^-$ as a factor in delaying star formation.
Furthermore, we neglect the photodissociation of H$^-$ due to radiation from local star formation for similar reasons, as explained in Johnson et al. (2007).

We have also neglected Ly$\alpha$ heating as a suppressant of minihalo star formation (see Chuzhoy \& Shapiro 2007; Ciardi \& Salvaterra 2007). For the Pop III star formation rates, and corresponding production rates of UV photons, found in our simulations, Ly$\alpha$ heating will not raise the temperature of the IGM appreciably, even for the maximal SFR considered here, ${\rm SFR}_{\rm III, max}$ (see Section 2.2).
Finally, we do not include a photoionizing background in our calculations. At these high redshifts, the IGM is still predominantly neutral. Photoionization will thus generally occur only locally, when isolated Pop~III stars form, an intermittent process which is not likely to dramatically affect the overall SFR (see e.g. Ahn \& Shapiro 2007; Johnson et al. 2007; Whalen et al. 2007; Wise \& Abel 2007a; but see Ricotti et al. 2002a,b). 

\section{Testing the Self-Regulated Model}
We have argued in Section 2.2 that the LW background and Pop III star formation rate at redshifts $z$ $\ga$ 15 are strongly coupled to one another, and that a self-consistent SFR is likely close to what is required to produce $J_{\rm LW, crit}$. To test this {\it Ansatz}, we have carried out a fiducial simulation with no LW feedback, and one with identical initial conditions, but with the critical LW background included. Comparing the resulting suppression of star formation in our simulations to the analytical ratio SFR$_{\rmn III, crit}/$SFR$_{\rmn III, max}$, we can roughly gauge the validity of our simple self-regulated model.

\subsection{Initial Conditions}
%VB: small changes here! "prohibitive" -> "much"
We initialize these cosmological simulations according to the $\Lambda$CDM model at $z$ = 100, adopting the cosmological parameters $\Omega_{m}=1 - \Omega_{\Lambda}=0.3$, 
$\Omega_{B}=0.045$, $h=0.7$, and $\sigma_{8}=0.9$, close to the values 
measured by {\it WMAP} in its first year (Spergel et al. 2003).  Here we use a periodic box with a comoving size $L$ = 500 $h^{-1}$ kpc, and particle numbers of $N_{\rm DM}$ = $N_{\rm SPH}$ = 145$^3$, where the SPH particle mass is $m_{\rm SPH}$ $\sim$ 710 ${\rm M}_{\odot}$.

As we have already noted in Section 2, conducting a sufficiently large cosmological simulation to thoroughly 
follow the build-up of the LW background is prohibitively expensive.  Our choice of initial conditions for the present simulation 
is thus purely a practical one, so that we may follow the evolution of a sufficient number of minihaloes in our cosmological box for our
purpose, which is to compare the suppression of the rate of collapse of gas in minihaloes with the suppression
of the star formation rate that is expected in our simple self-regulated model.  Because this simulation 
is not the large-scale cosmological simulation that is necessary to fully test the self-regulated model, all we can 
derive from this simulation is whether or not the delay in the collapse of the few ($\sim$ 20) minihaloes that form in our 
small simulation box is roughly consistent with the global average suppression of the star formation rate that we predict in the 
self-regulated model.  Therefore, the results that we derive from this test hold no statistical significance which is dependent
on the specific cosmological model that we use to initialize our simulation; we only seek to compare the isolated behavior of
the minihaloes in our box with the general behavior we expect in the self-regulated model.
A more detailed study of the self-consistent build-up of the LW background will be the subject of future work.        

The density above which resolution is lost in this simulation is $n_{\rm res}$ $\sim$ 20 cm$^{-3}$, as determined following Johnson et al. (2007).  Therefore, sink particles are formed in minihaloes which reach this density threshold, and we thereafter consider these minihaloes as hosting the formation of single Pop III stars.  While we are not able to resolve the process of star formation itself, we believe this is a reliable indicator of when star formation is possible, as gas collapsing in minihaloes will generally only reach these densities when cooling is efficient and the collapse has ceased to be adiabatic (e.g. Omukai et al. 2005; Yoshida et al. 2006; Johnson et al. 2007).

\begin{figure}
\vspace{2pt}
\epsfig{file=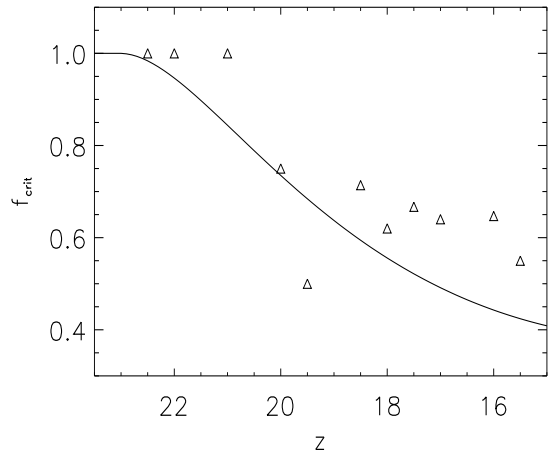,width=8.5cm,height=7.cm}
\caption{
The fraction of minihaloes which form stars when exposed to the critical LW background, $J_{\rmn LW, crit}$, normalized to the total number of stars formed absent a LW background, as a function of redshift.
{\it Solid line:} The fraction of Pop III stars expected to form in our analytical model for the critical star formation rate, SFR$_{\rmn III, crit}$, assuming the star formation efficiency $f_*$ shown in Fig. 1.  
{\it Open triangles:} The results from our simulation show the same general variation with $z$ as the analytically-derived behavior to within a factor of $\sim$ 2, with the points from our simulation likely being overestimates due to our choice of {\it WMAP} 
first year initial conditions for our simulation. 
This agreement supports the validity of our simple model for self-regulated Pop III star formation.}
\end{figure}

\subsection{Results}
In Fig.~5, we compare our analytical prediction, described in Section 2.3.1, to our simulation results. 
The line in Fig.~5 shows the analytically derived value for the fraction of 
stars, $f_{\rmn crit}(z)$, which form by a redshift $z$ in the self-regulated model out of the total number of minihaloes which could host star formation absent any negative LW feedback. We define this fraction as the ratio of the integrated SFR in the self-regulated model to the integrated SFR in the model for maximal star formation, described in Sections 2.3.1 and 2.2, respectively:

\begin{equation}
f_{\rmn crit} = \frac{\int_z^{\infty} {\rmn SFR}_{\rmn III, crit}(z')dz'}{\int_z^{\infty} {\rmn SFR}_{\rmn III, max}(z')dz'} \mbox{\ .} 
\end{equation}    
In this analytical treatment, we have assumed the value shown in Fig. 1 for the fraction of collapsed baryons in stars $f_*$, as a function of redshift.

The quantity $f_{\rm crit}$ is calculated as the number of sink particles that form by a redshift $z$ 
in the simulation including the critical LW background, $J_{\rm LW, crit}$, divided by the number 
that form in the simulation without any LW background included, and is plotted in Fig.~5. 
As expected, likely owing to the relatively high bias in our simulations initialized with {\it WMAP} first year parameters, the simulated values for $f_{\rm crit}$ are higher than those from our analytical derivation using {\it WMAP} third year parameters. Nonetheless, the agreement between our simple analytical model and the simulations is broadly consistent, given the simplicity of our model. We thus conclude that, as a rough estimate, a self-consistent value for the LW background at redshifts $z$ $\ga$ 15 is likely close the critical value, $J_{\rm LW, crit}$.  We emphasize, however, that the effect of the opacity to LW photons through relic H~II regions will be to lower $J_{\rm LW}$ and, simultaneously, to allow for a higher SFR than that in our simple self-regulated model shown in Fig.~2.    

\section{Galaxy Formation in the Early Universe}
Assuming that the photodissociation of H$_2$ molecules by a global LW background is the dominant factor in regulating Pop~III star formation in minihaloes at redshifts $z$ $\ga$ 15, we now employ our simple model for a self-regulated LW background to investigate the impact of Pop~III star formation on the assembly of a galaxy at $z$ $\ga$ 10.  We choose to define a galaxy as a star forming system hosted by a DM halo with a virial temperature of $\ga$ 10$^4$ K, corresponding to a halo with total mass $\ga$ 10$^8$ M$_{\odot}$ at $z$ $\ga$ 10 (e.g. Barkana \& Loeb 2001), as gas in such a system can cool via atomic hydrogen, can likely host continuous star formation, and can retain gas against being blown out by supernovae or by photoheating (e.g. Mori et al. 2002; Kitayama \& Yoshida 2005; Read et al. 2006).  
Here we report the results from our simulation of the formation of a galaxy which virializes at redshift $z$ $\sim$ 12, focusing on the modes and amount of star formation that may take place during its assembly.  Additional aspects of the formation of such a galaxy are discussed in a companion paper (Greif et al. 2008).

\subsection{Initial Conditions}
For our simulation of the assembly of a 'first galaxy', we have employed multi-grid initial conditions which offer higher resolution in the region where the galaxy forms (e.g. Kawata \& Gibson 2003; Li et al. 2007).
We again initialize the simulation according to the $\Lambda$CDM model at $z$ = 100, adopting the cosmological parameters $\Omega_{m}=1 - \Omega_{\Lambda}=0.3$, $\Omega_{B}=0.045$, $h=0.7$, and $\sigma_{8}=0.9$, close to the values 
measured by {\it WMAP} in its first year (Spergel et al. 2003). Here we use a periodic box with a comoving size $L$ = 1 $h^{-1}$ Mpc for the parent grid.  This simulation uses a number of particles $N_{\rm DM}$ = $N_{\rm SPH}$ = 1.05 $\times$ 10$^6$, where the SPH particle mass is $m_{\rm SPH}$ $\sim$  120 ${\rm M}_{\odot}$ in the region with the highest resolution. For further details on the technique employed to generate our multi-grid initial conditions, see the companion paper by Greif et al. (2008).
The density above which resolution is lost is $n_{\rm res}$ $\sim$ 10$^3$ cm$^{-3}$, as determined following Johnson et al. (2007). As described in Sections 3 and 4, sink particles are formed in minihaloes which reach this density threshold, and we thereafter consider these sinks to contain single Pop III stars. 

\begin{figure}
\vspace{2pt}
\epsfig{file=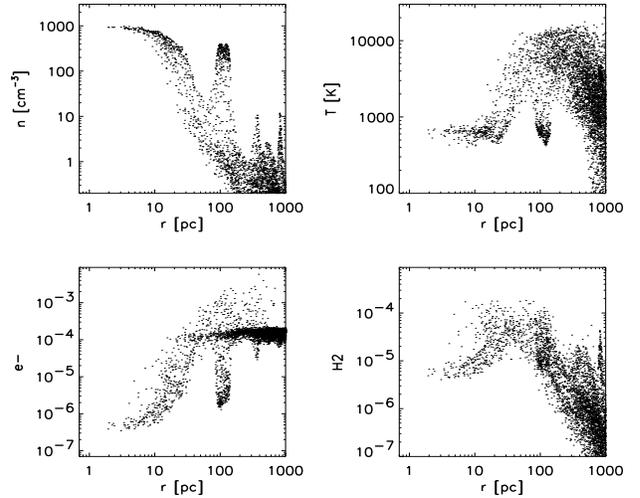,width=8.5cm,height=7.cm}
\caption{The properties of the primordial gas just before the formation of the first sink particle at $z$ = 16.3, as a function of distance from the sink: density (top-left), temperature (top-right), free electron fraction $f_{\rm e^-}$ (bottom-left), and H$_2$ fraction $f_{\rm H_2}$ (bottom-right). Note the enhanced free electron fraction in regions heated by the virialization of the host halo.  Also, note that a second star later forms at 100 pc from the first in this halo, the second halo having collapsed to densities above $n$ $\sim$ 500 cm$^{-3}$ when the first star forms, densities which yield the evolution of the gas in the second halo relatively impervious to the radiative feedback from the first star (Whalen et al. 2007; Susa \& Umemura 2006; Ahn \& Shapiro 2007).  Finally, note that the H$_2$ fraction is considerably lower than the values generally found for collapsing minihaloes in calculations not including a LW background.  This leads to a higher temperature of the dense gas in the core of the minihalo compared to the canonical case absent a LW background (e.g. Bromm et al. 2002). This may lead to a larger mass for the Pop III star that forms from this gas.  For reference, the resolution length in our simulation is $\sim$ 3 pc (physical).  
}
\end{figure}

\subsection{Formation of the First Star}
The first star forms at redshift $z$ $\sim$ 16.3 inside the minihalo which is the most massive 
progenitor of the halo eventually hosting the galaxy. The mass of the DM halo hosting the first 
star is $7 \times$ 10$^6$ M$_{\odot}$, while the virial temperature of the halo is 6 $\times$ 10$^3$~K.  This is comparable to the mass of the halo when star formation begins found in other studies for the same LW background and at similar redshift (O'Shea \& Norman 2008).  We note that this DM halo is approaching, but has not reached, virial equilibrium when the first star forms, having a gravitational potential energy of $\sim$ 1.2 $\times$ 10$^{52}$ erg and a kinetic energy of $\sim$ 1.1 $\times$ 10$^{52}$ erg.  

The properties of the primordial gas just before the star forms are different from the canonical case of 
Pop III star formation in the absence of a LW background, as shown in Fig. 6.  In the dense core of the halo, the H$_2$ abundance is $\la$ 10$^{-5}$, well 
below the canonical value of $\la$ 10$^{-3}$ at $n$ $\sim$ 10$^{3}$ cm$^{-3}$, owing to the photodissociation of H$_2$.  
In turn, the temperature of the gas in the core is $\ga$ 500 K, markedly higher than in the case without photodissociation, 
for which temperatures are generally $\sim$ 200 K.  We note that the central temperature that we find in the protostellar core 
is similar to what is found in the simulations of O'Shea \& Norman (2008) for a comparable value of the LW background flux; 
however, these authors find a higher central H$_2$ fraction of $\ga$ 10$^{-4}$, likely because they follow the 
collapse of the gas to higher densities, $n$ $>$ 10$^{4}$ cm$^{-3}$, at which H$_2$ formation may be faster.  The higher temperatures of the collapsing gas will likely translate into higher accretion rates onto the protostellar core than in the case without photodissociation (e.g. Omukai \& Palla 2001, 2003; Bromm \& Loeb 2004; O'Shea \& Norman 2008), and so possibly into a higher final mass for the star.  If the mass of the star is pushed above $\sim$ 260 M$_{\odot}$, or if it remains below $\sim$ 140 M$_{\odot}$, it may collapse directly to form a black hole at the end of its life without yielding a supernova explosion, thereby keeping its surroundings largely metal-free (e.g. Fryer et al. 2001; Heger et al. 2003). 

The mass of the DM halo hosting this star is roughly an order of magnitude below that required for the virial temperature of the halo to exceed the atomic-cooling threshold of 10$^{3.9}$ K, and is likewise well below the mass that we expect for the first galaxies.  However, star formation is still able to take place at this relatively early time, despite the effect of the global LW background. We conclude that it is likely that at least a single Pop III star forms by $z$ $\sim$ 15 in a DM minihalo which is the most massive progenitor of an atomic-cooling halo hosting a galaxy at $z$ $\ga$ 10.  The impact that such a star may have on the formation of the galaxy clearly depends sensitively on the final mass of this star.

\subsection{Local Stellar Feedback on Minihaloes}
We find that there is a second star-forming minihalo merely $\sim$ 100 pc from the site of the formation of the first star in our simulation, as can be seen in Fig. 6.  Because of this proximity, we expect that any radiative, mechanical, or chemical feedback from the first star may dramatically impact the evolution of this neighbouring halo. Numerous recent studies have discussed these feedback effects in detail (e.g. Ciardi \& Ferrara 2005), and we here draw on them to understand how this stellar feedback may impact the formation of a galaxy in the early Universe.

Regardless of whether the first star becomes a supernova or collapses directly to form a black hole, the neighbouring halo will be subject to the ionizing and molecule-dissociating flux from this star (e.g. Shapiro et al. 2004; Susa \& Umemura 2006; Ahn \& Shapiro 2007; Johnson et al. 2007; Yoshida et al. 2007; Whalen et al. 2007).  As the neighbouring halo will have already collapsed to densities $\ga$ 500 cm$^{-3}$ by the time the first star forms, the radiation from the first star is not likely to delay the collapse of the second halo significantly, as at these densities the core of the halo will not be photoevaporated and the molecules in the dense core may only be weakly dissociated (e.g. Susa 2007; Whalen et al. 2007).  Therefore, it appears that a second star will form in the neighbouring minihalo within $\la$ 10 Myr after the formation of the first star (see also Wise \& Abel 2007a).  If the first star collapses directly to form a black hole, the formation of the second star will thus not be significantly influenced by the first star.

The type of star that forms in the neighbouring halo is not entirely clear, however, if the first star becomes a supernova, owing to the mechanical and chemical feedback that this would entail.  As shown in Greif et al. (2007), the velocity of the shock produced by a Pop III supernova at a distance of 100 pc from the progenitor is ${v}_{\rm{sh}}$ $\la$ 100 km s$^{-1}$.  A shock propagating at such velocities can strongly ionize the primordial gas (e.g. Shull \& McKee 1979), thereby allowing the formation of a high fraction of molecules and enhancing the cooling properties of the gas (e.g. Shapiro \& Kang 1987; Mackey et al. 2003; Salvaterra et al. 2004; Machida et al. 2005).  In particular, high fractions of HD can be generated behind such strong shocks, allowing the post-shock gas to cool to the temperature of the cosmic microwave background (CMB) and perhaps to form metal-free stars with masses of the order of 10 M$_{\odot}$, so-called Pop III.2 stars. We note that for this type of Pop III.2 star formation to take place the pre-shock gas must be at sufficiently high densities, unlike in the supernova simulation of Greif et al. (2007), in which this mode of star formation did not occur in the low-density IGM (see also Bromm et al. 2003). However, the neighbouring minihalo in the present case is likely to be sufficiently dense to allow for star formation to take place in the post-shock gas.  We thus conclude that the halo neighbouring the first star may host a Pop III.2 star, in the case that the first star explodes as a supernova.

The mode of star formation in the neighbouring minihalo may change from Pop~III to Pop~II, if metal-rich supernova ejecta from the first star can efficiently mix with the primordial gas in the dense core of the minihalo (e.g. Bromm et al. 2001; Santoro \& Shull 2006; Schneider et al. 2006). We can estimate the efficiency of this mixing by applying the criterion for the operation of Kelvin-Helmholtz (KH) instabilities given by Murray et al. (1993) (see also Cen \& Riquelme 2008; Wyithe \& Cen 2007):
\begin{equation}
\frac{g D r}{2\pi v_{\rm{sh}}^2} \la 1\mbox{\ ,}
\label{mix}
\end{equation}
where $g$ is the gravitational acceleration at the outer radius of the dense core of the neighbouring minihalo, and $D$ the density ratio of gas in the halo compared to the dense shell. For the dense core of the neighbouring minihalo, we have a mass of the order of $10^{5}~M_{\odot}$ and a radius of $r \simeq 30~\rm{pc}$.  We find $D\ga 500$ and ${v}_{\rm{sh}}\la 100~\rm{km}~\rm{s}^{-1}$, using values for the density and velocity of the shocked gas from Greif et al. (2007).  The left-hand side of equation (18) is thus $\ga$ 10$^{-1}$. Following the results of Cen \& Riquelme (2008), this suggests that the dense gas in the core of the neighbouring minihalo will remain largely pristine and stable for many dynamical times, while the outskirts of the halo may become more mixed.  It therefore appears unlikely that the neighbouring halo will give rise to Pop II stars, but will instead probably host Pop~III star formation.  This will have implications for the nature of the galaxy into which these minihaloes are finally incorporated.

\begin{figure}
\vspace{2pt}
\epsfig{file=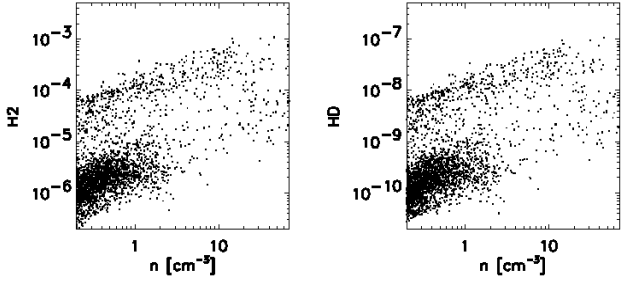,width=8.5cm,height=4cm}
\caption{ The fraction of molecules in the primordial gas of the protogalaxy at $z$ $\sim$ 12.5, as functions of density: H$_2$ fraction $f_{\rm H_2}$ (left), and HD fraction $f_{\rm HD}$ (right).  The free electron fraction greatly exceeds the primordial value of $f_{\rm e^-}$ $\sim$ 10$^{-4}$ through collisional ionization in virialization shocks, leading to enhanced fractions of both H$_2$ and HD, despite the presence of the LW background radiation field.  Accounting for self-shielding, the molecule fraction remains elevated at high densities $n$ > 10$^2$ cm$^{-3}$.  In the core of the halo, the HD fraction is thus expected to reach values large enough to allow the formation of Pop III.2 stars (see Greif et al. 2008).}
\end{figure}

\subsection{Self-shielding and Molecule Formation}
%VB: a few changes/questions in this subsection!
While minihaloes are, in general, optically thin to LW photons, the greater gas mass in a halo hosting the formation of a first galaxy can 
lead to a large column density of H$_2$ and so to self-shielding of the molecules in the central regions of the halo (e.g. Draine \& Bertoldi 1996).  
Taking this into account, we would like to estimate the impact of the LW background on the molecule fraction in the central regions of a first galaxy.  
Assuming that H$_2$ formation takes place through the reaction sequence 

\begin{equation}
{\rmn e^{-}} +{\rmn  H} \to {\rmn H^{-}} + h\nu \mbox{\ ,}
\end{equation} 
       
\begin{equation}
{\rmn H^{-}} +{\rmn  H} \to {\rmn H_{2}} + {\rmn e^{-}} \mbox{\ ,}
\end{equation} 
we can estimate the formation rate of H$_2$ as the gas collapses to high density.  In our simulation, we find that the electron fraction drops due to 
recombinations approximately as

\begin{equation}
f_{\rm e} \sim 10^{-4}\left(\frac{n_{\rm H}}{10^2 {\rm cm}^{-3}}\right)^{-2}  \mbox{\ ,}
\end{equation} 
where $n_{\rm H}$ is the number density of hydrogen nuclei in the gas.  Following Bromm et al. (2002), this results in a formation 
rate of H$_2$ at densities $n_{\rm H}$ $\ga$ 10$^2$ cm$^{-3}$ of

\begin{equation}
\left(\frac{df_{\rm H_2}}{dt}\right)_{\rm form} \sim k_{\rm H-} n_{\rm H} f_{\rm e} \sim k_{\rm H-} n_{\rm H}^{-1} \mbox{\ ,} 
\end{equation}
where $k_{\rm H-}$ $\sim$ 10$^{-15}$ cm$^3$ s$^{-1}$ is the rate coefficient for reaction (19).   

We can next estimate the photodissociation rate of H$_2$, including the effect of self-shielding, following Bromm \& Loeb (2003a).  
Accordingly, we estimate the column density of H$_2$, using only local quantities, as $N_{\rm H_2}$ $\sim$ 
0.1$f_{\rm H_2}$$n_{\rm H}$$L_{\rm char}$, where the local characteristic length is given by

\begin{equation}
L_{\rm char} = \left(\frac{3 X M_{\rm b}}{4 \pi m_{\rm H} n_{\rm H}} \right)^{1/3} \mbox{\ .}
\end{equation}
Here M$_{\rm b}$ $\sim$ 10$^7$ $M_{\odot}$ is the total baryonic mass in the halo, $X=0.76$ is the mass fraction of hydrogen, and $m_{\rm H}$ is the mass of 
the hydrogen nucleus.  From our simulation of the formation of a first galaxy, which does not include the effects of self-shielding, 
we find a minimum average molecule fraction of $f_{\rm H_2}$ $\sim$ 10$^{-5}$ at densities $n_{\rm H}$ $\ga$ 10$^2$ cm$^{-3}$, as shown in Fig. 6, when 
$J_{\rm LW}$ = $J_{\rm LW, crit}$.  The photodissociation rate will be reduced due to self-shielding by a fraction 

\begin{equation}
f_{\rm sh} \sim \left(\frac{N_{\rm H_2}}{10^{14} {\rm cm}^{-2}}\right)^{-\frac{3}{4}} \mbox{\,}
\end{equation}
when $N_{\rm H_2}$ $>$ 10$^{14}$ cm$^{-2}$ (Draine \& Bertoldi 1996).  We thus find a minimum photodissociation rate of 

\begin{equation}
\left(\frac{df_{\rm H_2}}{dt}\right)_{\rm diss} \sim 3 \times 10^{-17} n_{\rm H}^{-1/2} f_{\rm H_2}^{1/4} J_{\rm LW} \mbox{\ .}
\end{equation}          
Balancing the formation rate and dissociation rates for H$_2$, given by equations (22) and (25), respectively, we can solve for the density at which the dissociation rate 
becomes higher than the formation rate, marking the point in the collapse of the gas where H$_2$ dissociation begins to substantially suppress the formation of H$_2$.  
We find this density to be
%VB: I think, the equation below is not quite correct. The dependence on f_H2 should be +1/2, I think.
%    I ve changed this. Please check! Also, when plugging in the numbers, I get a smaller n_diss.
%    Please carefully check!

\begin{equation}
n_{\rm diss} \sim 10^8 {\rm cm}^{-3} \left(\frac{J_{\rm LW}}{0.04}\right)^{-2} \left(\frac{f_{\rm H_2}}{10^{-5}}\right)^{-1/2} \mbox{\ .}
\end{equation}
For the values of J$_{\rm LW}$ = 0.04 and $f_{\rm H_2}$ $\sim$ 10$^{-5}$ in our simulation, we thus find that H$_2$ photodissociation becomes important only at very high densities.  At densities below $n_{\rm diss}$, the formation rate of H$_2$ is higher than the photodissociation rate, owing to the higher abundance of free electrons, which catalyze the formation of H$_2$ according to equations (19) and (20).  As the density increases, the recombination of free electrons deprives the gas of this catalyst, and the formation rate of H$_2$ molecules thus drops, becoming comparable to the photodissociation rate when the gas has collapsed to a density $n_{\rm diss}$.  
We note that photodissociation may not become important at all in the evolution of the gas, if $n_{\rm diss}$ $\ga$ 10$^8$ cm$^{-3}$, at which densities three-body reactions can efficiently produce H$_2$ (see e.g. Glover 2005).  Furthermore, the photodissociation of H$_2$ is not likely to affect the evolution of the gas for cases in which the photodissociation timescale is longer than the free-fall timescale, as the gas may then cool and collapse to form a star before a significant portion of the H$_2$ molecules are photodissociated.

Along with the likely elevated fraction of H$_2$, owing to ionization of the gas in the virialization of the galaxy, 
Fig. 7 shows that a high fraction of HD is also generated.  We expect that, due to the substantial self-shielding of the gas in the central regions of the assembling galaxy, 
the high HD fraction will persist in the gas as it collapses to high density, likely leading to the formation of Pop~III.2 stars (see Greif et al. 2008).

\begin{figure*}
\includegraphics[width=7.in]{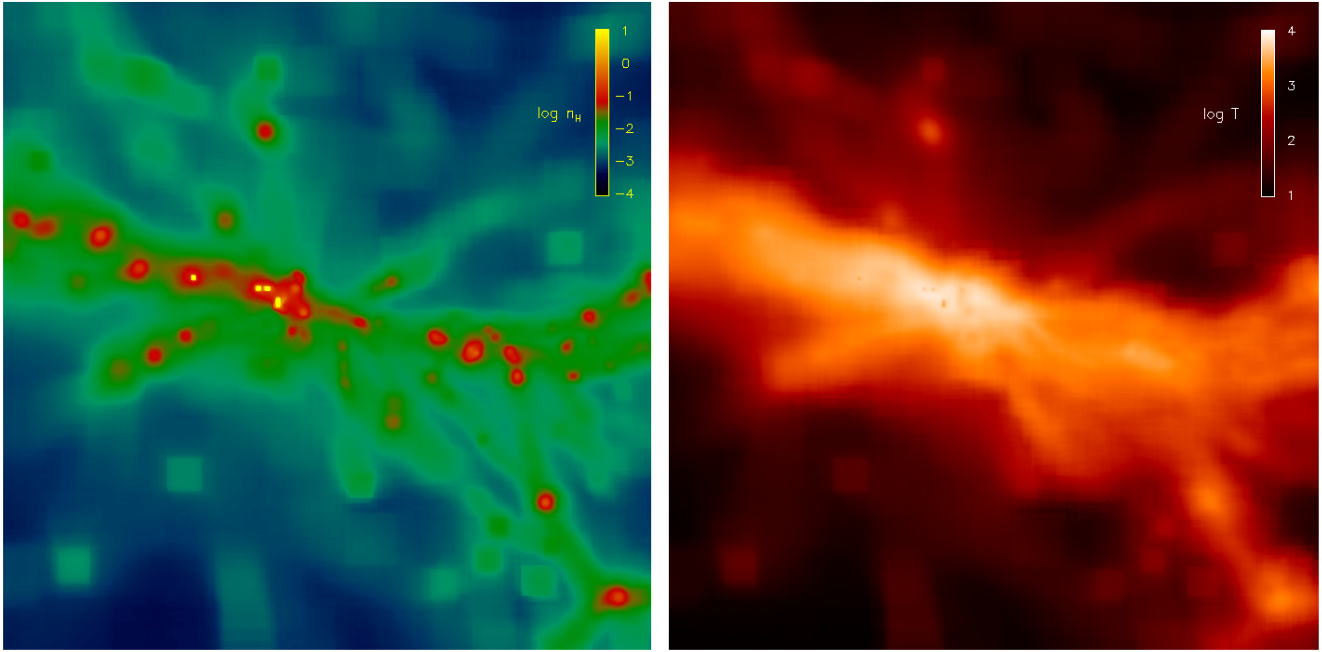}

\caption{The hydrogen number density and temperature of the gas in the region of the forming galaxy at $z$ $\sim$ 12.5.  
The panels show the inner $\sim $ 10.6 kpc (physical) of our cosmological box.   The cluster of minihaloes harboring dense 
gas just left of the center in each panel is the site of the formation of the two Pop III stars which are able to form 
in our simulation including the effects of the self-regulated LW background.  The remaining minihaloes are not able to form 
stars by this redshift, largely due to the photodissociation of the coolant H$_2$.  The main progenitor DM halo, which hosts 
the first star at $z$ $\sim$ 16, by $z$ $\sim$ 12.5 has accumulated a mass of 9 $\times$ 10$^7$ M$_{\odot}$ through 
mergers and accretion.  Note that the gas in this halo has been heated to temperatures above 10$^4$ K,  leading to a high 
free electron fraction and high molecule fractions in the collapsing gas (see Fig. 7). The high HD fraction that is 
generated likely leads to the formation of Pop III.2 stars in this system. 
}
\end{figure*}

\section{Implications for the Assembly of the First Galaxies}
In the preceding sections we have found that, owing in part to the self-regulation of the LW background, the Pop III star formation rate in minihaloes in the early Universe remains substantial at redshifts $z$ $\ga$ 15.  In particular, we have found that the LW background is likely too weak to completely suppress star formation in the progenitor DM minihaloes that merge to from the first galaxies at $z$ $\ga$ 10. 
Here we discuss various implications of these results for the nature of the first galaxies.  

\subsection{The Protogalactic Star Formation Rate}
While, in principle, Pop III star formation can be suppressed in minihaloes by a global LW background, we have shown that with a simple model for the self-consistent build-up of the LW background Pop III star formation is likely to take place in the assembly of a galaxy at $z$ $\ga$ 10 (see Fig.~8).  Noting that the LW background that we used in our simulation did not take into account the effect of the opacity to LW photons through relic H~II regions in the early Universe (see Section 2.3), the negative feedback from the global LW background may be even weaker than we found in our simulation.  This, in turn, implies that the number of Pop III stars forming in protogalaxies may be higher than we find from our simulation.  A total of two minihaloes collapsed to form Pop III stars in our simulation employing the self-regulated model for the build-up of the LW background, allowing us to conclude that of the order of a few Pop III stars are likely able to form during the assembly of a galaxy at high redshift.  

  Additionally, we have found that conditions are prime for the formation of Pop III.2 stars, owing to the high HD fraction that is generated in the virialization shocks in such systems. Because the Pop~III.2 star formation mode in atomic-cooling haloes has not been studied with simulations to the same extent as the `classical', minihalo (Pop~III.1) case, it is not yet possible to predict the resulting star formation rate and efficiency (but see Greif \& Bromm 2006). However, we expect that, with 10$^7$ M$_{\odot}$ of gas collapsing into such a system, Pop~III.2 stars may form in clusters, given that these stars are expected to have masses of the order of 10 M$_{\odot}$, compared to the 100 M$_{\odot}$ predicted for Pop~III.1 stars formed in minihaloes. The stellar feedback is then expected to be less severe, possibly enabling cluster formation.

\subsection{Supernova Feedback and Metallicity}
Each of the two star-forming haloes in our simulation is located within $\sim$ 100 pc of the center of the main progenitor of the protogalaxy, allowing each star to affect the inner region of the galaxy through either radiative output or by exploding as supernovae. We can estimate the degree of metal enrichment that may take place should each of these stars explode as pair-instability supernova (PISN), expected to be a common fate for massive Pop~III stars (e.g. Heger et al. 2003), as follows. For a 10$^8$ M$_{\odot}$ DM halo, we expect $\sim$ 10$^7$ M$_{\odot}$ of gas to collapse into the virialized system. With $\sim$ 100 M$_{\odot}$ in metals being expelled into the medium from two PISNe, we thus expect to have a mean metallicity in this system of $\sim$ 10$^{-3}$ $Z_{\odot}$. 

Interestingly, this is above the value for the critical metallicity for Pop II star formation, $Z_{\rm crit}$ $\sim$ 10$^{-3.5}$ $Z_{\odot}$, as given by Bromm \& Loeb (2003b) (see also Bromm et al. 2001a; Frebel et al. 2007; Smith \& Sigurdsson 2007). This metallicity estimate assumes very efficient mixing of the metals with the virialized primordial gas. In reality, there will likely be pockets of less enriched gas in the galaxy, suggesting that in the first galaxies there may be regions in which Pop~II star formation is enabled through metal cooling, as well as other, more pristine regions in which a top-heavy IMF persists (see Jimenez \& Haiman 2006).  Considering that the SFR may in fact be larger than what we found in our simulation, owing to weaker LW feedback, we expect that the mean metallicity in the first protogalaxies may easily exceed the critical metallicity necessary for low-mass Pop~II star formation. If the mixing of these metals is efficient, the first galaxies may already host primarily Pop~II star formation with a preponderance of low-mass stars, provided that the mass of the stars formed in these systems is not dictated instead by other effects, e.g. the inability of the gas to cool below the temperature of the CMB, $T_{\rm CMB} = 2.7{\rm \ K\,}(1 + z)$ (e.g. Larson 1998, 2005).
 
However, the details of the transition from a top-heavy Pop III IMF to a low-mass Pop II IMF are not at present well understood, as it remains to understand the process by which the first metals, ejected into a hot expanding supernova remnant, are ultimately reincorporated into dense gas from which stars can form.  The cooling time of the hot metal-rich ejecta from an energetic Pop III supernova is of the order of 10$^8$ yr (Greif et al. 2007).  This is ample time for multiple Pop III stars to form from cold pockets of primordial gas and explode as supernovae, allowing to enrich the gas with the yields from multiple supernovae before it re-collapses to form the first Pop II stars.  Such a mixing of SN yields would dilute any pure PISN signature from a single massive Pop III star, and may explain the lack of observed metal-poor stars exhibiting a clear PISN signature (see Karlsson et al. 2008). An additional complication might
arise from the presence of cosmic rays (CRs) produced by the first supernovae.
Heating and ionization due to CRs could have an important effect on the
early star formation process (e.g. Jasche et al. 2007; Stacy \& Bromm 2007), although these authors find that the main effect of cosmic rays is to enhance molecule formation in the
central dense regions of minihaloes, which only strengthens our main conclusion that Pop~III star formation in minihaloes is not easily suppressed during the assembly of the first galaxies.

\subsection{Black Hole Formation}
A fraction of the first stars likely collapsed directly to form black holes, thereby locking up the majority of the metals produced in their cores (see e.g. Madau \& Rees 2001; Fryer et al. 2001; Heger et al. 2003).  In a given protogalactic system, there is thus a probability that the primordial gas will not be enriched by supernovae even if stars are able to form within this system.  Without knowing the IMF of Pop III stars, however, we can only guess at what may be the fraction of Pop III stars which collapse to form black holes without ejecting appreciable amounts of metals.  Heger et al. (2003) give the mass range for Pop III stars that collapse directly to black holes as $\sim$ 40 to $\sim$ 140 M$_{\odot}$ and $\ga$ 260 M$_{\odot}$, while that over which a PISN takes place is given as $\sim$ 140 to $\sim$ 260 M$_{\odot}$, for stellar models which neglect the effects of rotation.  

Given that Pop III stars formed in minihaloes are characterized by a top-heavy IMF with typical masses of the order of 100 M$_{\odot}$, it seems reasonable to assume that the probability for a Pop III star to collapse directly into a black hole, and thus locking up its metal-rich core, is of the order of 0.5. We note that this is consistent with the constraints on the fraction of PISNe from Pop III stars reported by Karlsson et al. (2008).  If we then take it that our simulation of the assembly of a $\sim$ 10$^8$ M$_{\odot}$ galaxy is typical and use the fact that we find of the order of a few stars forming during this assembly process, we obtain a probability of the order of $\la$ 0.1 that the first galaxy may be metal-free at the time of its formation.  While this is only a rough estimate, the fact that we find little chance that star formation can be completely suppressed during the assembly of a galaxy at redshifts $z$ $\ga$ 10 suggests that metal-free galaxies are probably a minority population of the first galaxies, with a majority of galaxies already hosting some amount of Pop II star formation.  We emphasize, however, that Pop III star formation may continue to occur, even after the first supernovae, in regions of a galaxy which are not enriched with metals or in which the mixing of metal-rich supernova ejecta with primordial gas is inefficient.  

Finally, as discussed in Section 5.2, given that the LW background can suppress H$_2$ cooling even in gas that does collapse to form Pop III stars, the high accretion rates in the formation of these objects may allow for a large fraction of such Pop III stars to have masses above $\sim$ 260 M$_{\odot}$, allowing them to directly collapse into black holes and thus eject few metals.  The LW background radiation may therefore act to keep galaxies metal-free even in cases where Pop III star formation is not prevented.

\section {Summary and Conclusions}
We have investigated the impact of the H$_2$-dissociating LW background radiation on the formation of Pop~III stars during the assembly of the first galaxies at $z$ $\ga$ 12. To this end, we have constructed simple models of the build-up of this radiation field which are self-consistent, 
accounting for the coupling of the star formation rate to the generation of the LW background.  
We have motivated and, to the extent possible, checked the self-consistency of a model in which 
the LW background does not exceed the critical value $J_{\rm LW, crit}$ $\sim$ 0.04 at redshifts 
$z$ $\ga$ 15.  Furthermore, we have elucidated the effect of a high opacity to LW photons 
through the first relic H~II regions on the LW background, finding that the LW background flux may take values $\la$ $J_{\rm LW, crit}$ at 15 $\la$ $z$ $\la$ 18, if the cosmological average optical depth to LW photons becomes sufficiently high.  While our models are idealized, they are informed by detailed cosmological simulations of Pop III star formation and 
are consistent with previous results.  We suggest that the actual LW background most likely assumes values intermediate between $J_{\rm LW, shield}$ and $J_{\rm LW, crit}$.  However, detailed large scale simulations tracking star formation and accounting for the opacity of the IGM to LW photons will be required to better understand the true nature of the LW background at high redshifts.  In addition, observations of the cosmic near-infrared background could conceivably be used to place conservative upper limits on the level of the LW background at high redshifts, as LW photons would be reprocessed into Ly$\alpha$ photons as they are redshifted with the expansion of the Universe, thus allowing them to contribute to the total Ly$\alpha$ flux emitted at high $z$ that may be observed today (e.g. Santos, Bromm \& Kamionkowski 2002; Fernandez \& Komatsu 2006).  
Taking into account the LW background, we find that the comoving rate of Pop III star formation in minihaloes at 25 $\ga$ $z$ $\ga$ 15 is likely $\la$ 10$^{-4}$ M$_{\odot}$ yr$^{-1}$ Mpc$^{-3}$, with the upper limit corresponding to the case of a maximal opacity to LW photons through relic H~II regions at these redshifts. Overall, we find that the global Pop III SFR is likely decreased from the maximum SFR rate possible by a factor of, at most, $\sim 3$ due to the LW background at $z$ $\ga$ 15.

We have simulated the assembly of a galaxy at $z$ $\ga$ 12, assuming the critical LW background in our self-regulated model of global Pop III star formation. We find that of the order of a few Pop III stars are likely to form in the minihaloes that later merge to form the galaxy.  Due to the chemical feedback that accompanies the explosions of Pop III stars as supernovae, possibly as PISNe, we conclude that most galaxies may already be enriched to a metallicity of the order of $\sim$ 10$^{-3}$ $Z_{\odot}$ when they are formed.  If this is the case, then Pop II star formation may be widespread in these systems even at these early times.  Alternatively, if the IMF of Pop III stars is such that a large fraction of them collapse directly to form black holes, thereby locking up the metals produced in their cores, then the interstellar medium of the first galaxies is more likely to remain metal-free.  
We find that a high fraction of the coolant HD is produced in the primordial gas during the virialization of the galaxy, despite the presence of the LW background, which likely leads to the formation of metal-free stars with masses of the order of 10 M$_{\odot}$ in these systems.  If metal-free galaxies do exist, they may therefore be dominated by such a population of stars, perhaps the first stellar clusters (see also Clark et al. 2008).  More detailed simulations tracking the collapse and fragmentation of the gas, along with feedback from Pop III stars, will be necessary in order to test this scenario.  

The character of star formation clearly becomes more complicated in the first galaxies compared to the canonical picture of Pop~III star formation in minihaloes, with the chemical feedback from the first stars, turbulence driven by rapid mergers and accretion of gas from the IGM (see Wise \& Abel 2007c; Greif et al. 2008), HD cooling, CR heating and possibly magnetic fields all playing important roles. Future missions, such as the {\it James Webb Space Telescope} may thus be expected to observe some variety in the first galaxies. The majority is likely already enriched with heavy elements, hosting Pop~II star formation, while a small fraction might still be largely metal-free, possibly hosting clusters of Pop~III.2 stars.
The veil is about to be lifted, giving us access to the time when galaxy
formation first began.

\section*{Acknowledgments}
We would like to thank Simon Glover for valuable comments that have improved
the presentation of this work.
VB acknowledges support from NSF grant AST-0708795 and NASA {\it Swift}
grant NNX07AJ636. The simulations presented here were carried out at
the Texas Advanced Computing Center (TACC).

\end{document}